\documentclass[12pt]{article}
\usepackage{epsf,epsfig}
\begin{document}
\title{Elastic Properties of Nematic Liquid Crystals Formed by Living and Migrating Cells }
\author{Ralf Kemkemer$^*$, Dieter Kling$^*$, Dieter Kaufmann$^{**}$ and Hans Gruler$^*$\\[20pt]
 Department of bioPhysics$^*$ and Human Genetics$^{**}$\\
University of Ulm\\
D 89069 Ulm, Germany \\[20pt]
e-mail: hans.gruler@physik.uni-ulm.de\\[20pt]
PACS number(s): 87.10.+e, 87.22.Nf, 82.70.-y \\[50pt] }
\date{\today}
\maketitle
\thispagestyle{empty}

\newpage
\begin{abstract}
In culture migrating and interacting amoeboid cells can form nematic liquid crystal phases. A polar nematic liquid crystal is formed if the interaction has a polar symmetry. One type of white blood cells (granulocytes) form clusters where the cells are oriented towards the center. The core of such an orientational defect (disclination) is either a granulocyte forced to be in an isotropic state or another cell type like a monocyte. An apolar nematic liquid crystal is formed if the interaction has an apolar symmetry. Different cell types like human melanocytes (=pigment cells of the skin), human fibroblasts (=connective tissue cells), human osteoblasts (=bone cells), human adipocytes (= fat cells) etc., form an apolar nematic liquid crystal. The orientational elastic energy is derived and the orientational defects (disclination) of nematic liquid crystals are investigated. The existence of half-numbered disclinations show that the nematic phase has an apolar symmetry. The density- and order parameter dependence of the orientational elastic constants and their absolute values are estimated. From the defect structure, one finds that the splay elastic constant is smaller than the bend elastic constant (melanocytes). The core of a disclination is either a cell free space or occupied by non oriented cells (isotropic phase) or occupied by a cell with a different symmetry or occupied by another cell type.      
\end{abstract}
\newpage 

\section*{Introduction}

Investigations of the structure and physical properties of liquid crystals have greatly increased in the last decades and are now an integral part of solid-state physics. Certain organic materials do not show a single transition from solid to liquid, but rather a cascade of transitions involving new phases. The mechanical properties and the symmetry properties of these phases are intermediate between those of a liquid and those of a crystal \cite{degennes}. 

Depending upon the nature of the building blocks and upon external parameters (temperature, solvents, etc.) a wide variety of phenomena and transitions amongst liquid crystals are observed. To generate a liquid crystal one must use anisotropic objects, like elongated molecules. Several ways are known to achieve this: with small molecules; with long helical rods that either occur in nature or can be made artificially; with polymers; with more complex structure units like the capside of a virus \cite{alberts} that are associated structures of molecules and ions; likewise, with even more complex structure units, like amoeboid cells which are complex structures far from thermodynamic equilibrium and act as a machine \cite{PMNcluster,dewald}. Here, the orientational elastic properties of liquid crystal formed by migrating and living cells are investigated.  

Thermotropic nematic liquid crystals can be formed by small elongated molecules like p-azoxyanisole (PAA) or N-(p-methoxy\-benzylidene)-p-butylaniline (MBBA) which can be regarded as rigid rods ($ \approx 2$ nm length and $\approx 0.5$ nm width). At low temperature one observes a crystalline phase. Heating up the sample, a transition to a nematic phase is observed. With further heating the transition to an isotropic liquid is observed. 

Lyotropic nematic liquid crystal can be formed by long elongated molecules in suitable solvents. The anisotropic building blocks can have a rod-like conformation like (i) synthetic polypeptides ($\approx 30 nm$ length and $\approx 2$ nm width) and (ii) tobacco mosaic virus  ($\approx 300 - 1000$ nm length and $\approx 20$ nm width). At low concentration one observes an isotropic liquid but a transition to a nematic phase is observed by increasing the concentration of the elongated building blocks.   

The fundamental property of a nematic liquid crystal, which makes it different from an isotropic liquid and similar to a solid, is the presence of an orientational degree of freedom which is characterized by a macroscopic spatial ordering of the long axis of molecules. The statistical mechanics for predicting the macroscopic spatial ordering is difficult but can be approximated by applying mean-field calculations. (i) The Onsager approach starts with anisotropic steric repulsion. The free energy is minimum for oriented rods (width $<<$ length): they show an orientational order of a nematic phase at high rod density. But, at low rod density, the suspension is an isotropic liquid, as expected. (ii) The Maier-Saupe theory approaches the intermolecular interactions by a mean-field which is proportional to the order parameter of the nematic phase. The minimum of the free enthalpy yields the equilibrium state. At high temperature an isotropic liquid without orientational order is predicted, but a nematic liquid crystal with orientational order can occur at low temperatures. 

\begin{figure}
\epsfxsize=7cm
\epsfbox{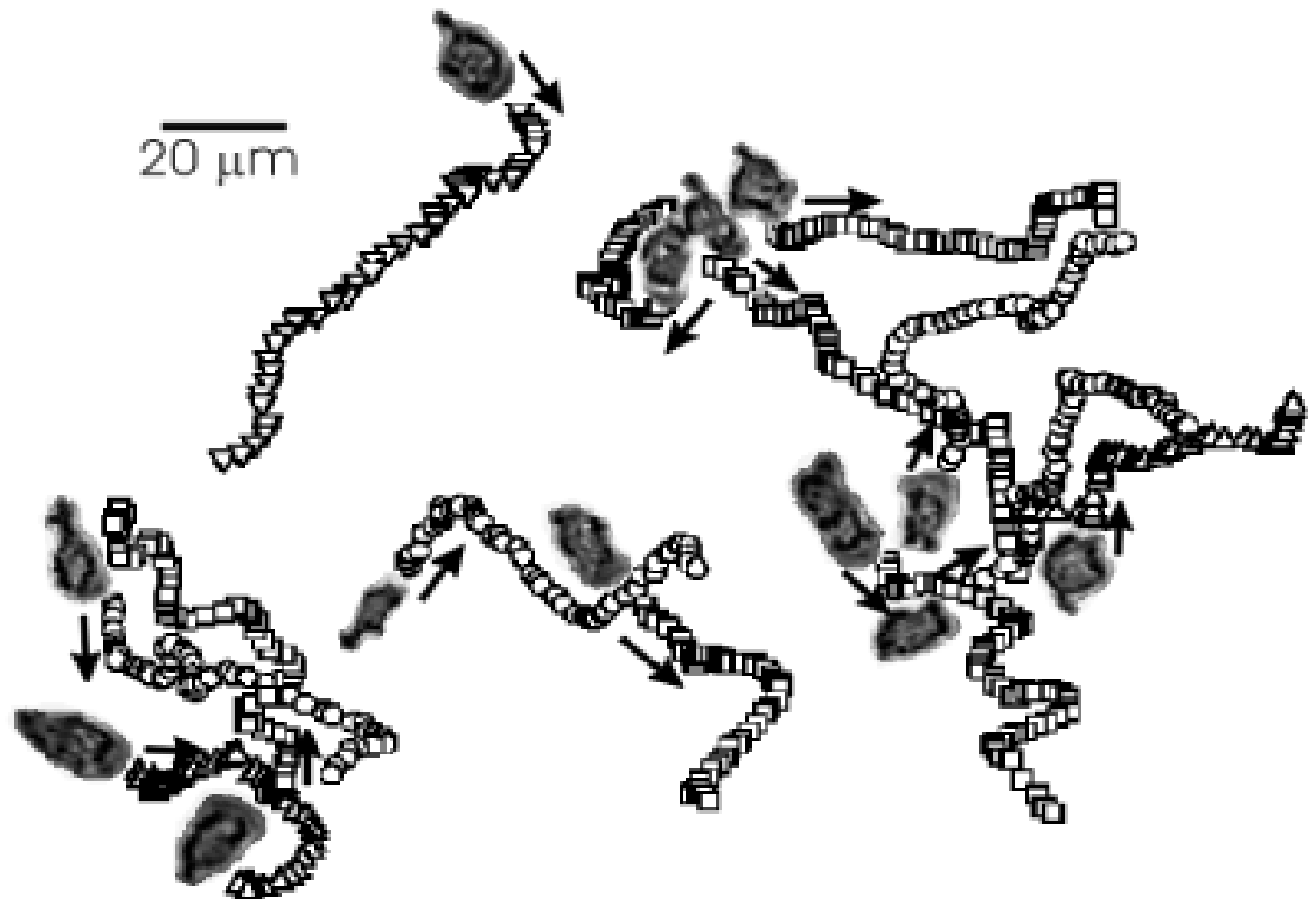}
\epsfxsize=7cm
\epsfbox{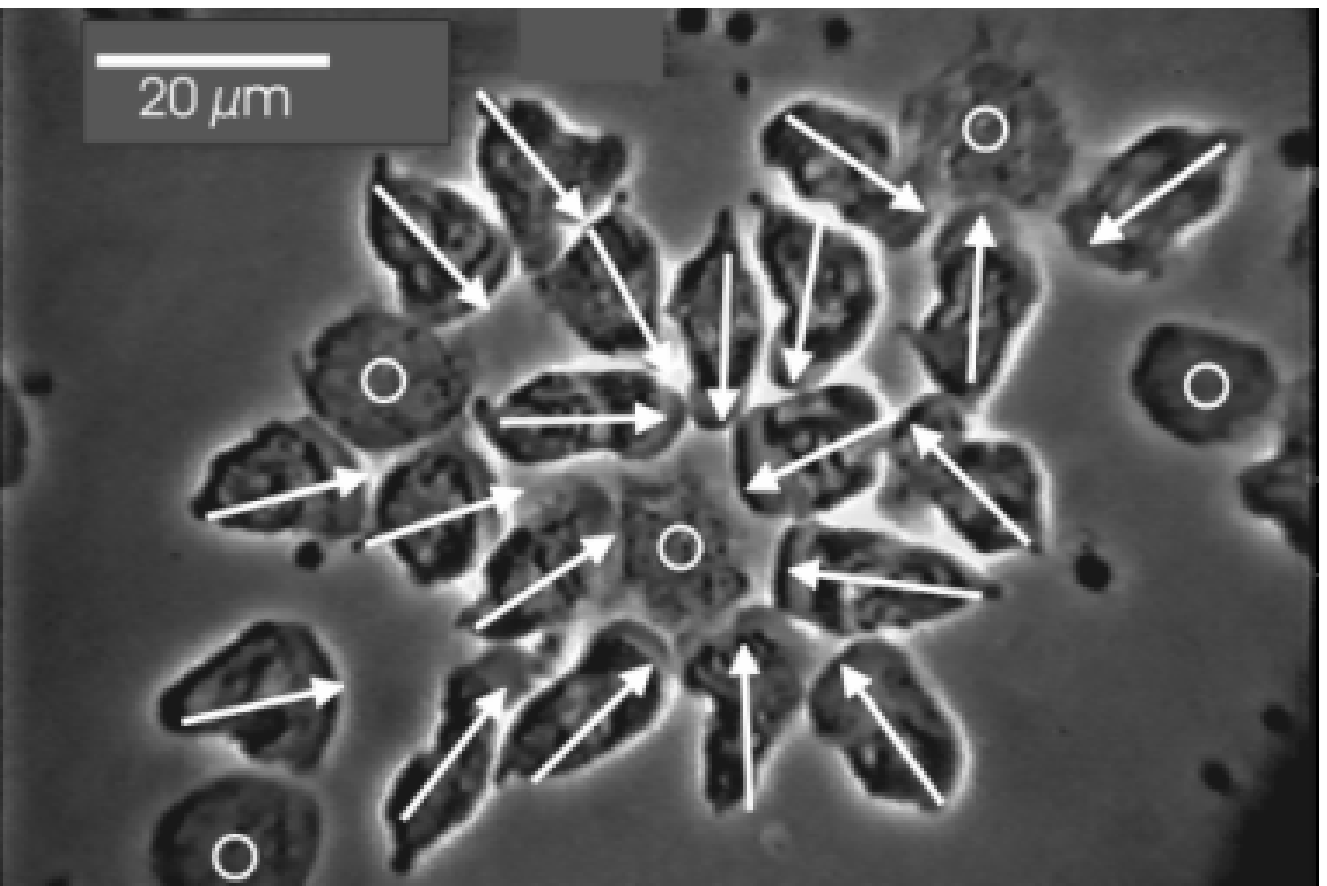}
\caption{\label{Cluster1} \it {
Human leukocytes exposed to blood plasma are migrating on a glass surface. The gap between the two glass slides is $ \approx 20 \mu$m. The fringe at the front of the cells is the amoeboid motor which pulls the rest of the cell. a) Single cell migration is observed at normal calcium concentration (2.5 mM). 70 pictures were digitized in a time period of 390 s. The trajectories of the single cells are shown. b) The single migrating cells start to form a cluster of dense packed cells if the calcium of the blood plasma is bound by EDTA ($<5 \mu$M calcium). The direction of migrating granulocytes is marked by an arrow. The immobile monocytes are marked by a circle. The monocycte close to the center of the picture acts as nucleation center of the migrating granulocytes. The polar order of the migrating cells in the cluster is very high (0.82). The monocyte at the upper right corner starts to act as an additional nucleation center.
}}
\end{figure}

The above mentioned nematic liquid crystals are described by equilibrium thermodynamic. This holds true even for tobacco mosaic virus since only the anisotropic shape of the viruses is considered. The intracellular metabolism can be neglected since the virus does not have the ability of self-motion.  Thus, the thermal motion is the stochastic source for the molecular dynamics. However, liquid crystal phases can also be formed if the stochastic source is created by energy consuming processes. A typical example are migrating biological cells like white blood cells. In the absence of a directing extracellular signal, the single cells perform a random walk due to the existence of stochastic processes in the intracellular signal transduction chain. The migrating cells can form liquid crystal phases if each cell transmits extracellular signals that are received by other cells. The symmetry of the cell-cell interaction is important for the formed type of liquid crystal phase: 

(i) A polar nematic liquid crystal is expected if the cell-cell interaction is polar. This means that one cell transmits a signal which attracts another cell. The cellular response is directed migration. For example: (a) Single slime mould cells migrate as amoeboid cells on a surface and search for bacteria as food. In the case of low bacteria concentration, the intracellular metabolism of the slime mould cells alters in such a way that cyclic AMP is emitted as an extracellular signal. Other slime mould cells can receive this chemical signal and react accordingly to the received signal with directed migration. At high enough cell density, the cells form a cluster where each cell tries to move towards the center of the cluster \cite{slime}. (b) A similar process is observed with migrating white blood cells \cite{PMNcluster}. If human leukocytes (granulocytes) are expose to blood plasma they then migrate as single amoeboid cells on a surface as shown in Fig. \ref{Cluster1}a. But if the calcium concentration is lowered, e.g. by adding the chelator EDTA (ethylene diamine tetraacetic acid, $pK_{Ca} = 10.7$) or  EGTA (ethyleneglycol-(aminoethyl ether) tetraacedic acid, $pK_{Ca} = 11$) to the plasma, then the migrating granulocytes attract each other and form a cluster, as shown in Fig. \ref{Cluster1}b. The emitted molecules (chemical signal) is still unknown. The process is reversible since the clusters disappear if the calcium concentration is increased. In both examples, the cells in a cluster form a polar nematic liquid crystal because (i) there is order in the cell orientation, $\langle cos \varphi \rangle$, induced by the directed migration and (ii) no order in the center of mass of the cells. The angle, $\varphi$, is the difference between the actual migration direction and the direction towards the center of the cluster. The cells in the cluster of Fig. \ref{Cluster1}b  have a polar order of 0.82. A singularity in the cell orientation is expected in the center of the cluster. A common observation is that the cell in the center has a spherical shape and the surrounding cells have a polar shape. Human monocytes (another type of leukocytes) are immobile on a glass surface. If a mixture of granulocytes and monocytes is investigated, the immobile monocytes act as nucleation center as shown in Fig. \ref{Cluster1}b. The polar nematic liquid crystal formed by living and migrating cells is a system far from thermodynamic equilibrium. 

One can ask which temperature corresponds to the random movement of the cells? This question can be answered by using the Einstein relation which connects the random movement of an inert particle described by the diffusion coefficient D, the mobility R and the thermal energy $k_B$T: $D R = k_B T$. The mobility of a spherical particle with radius r in a viscous medium with viscosity $\eta$ is given by Stokes: $R = 6 \pi \eta r$. The random movement  of a migration granulocyte can be quantified by measuring the mean-squared displacement as a function of time. The diffusion coefficient, $D$, is one fitting parameter ($D = 200 - 1000 \mu m^2/min$). The calculated temperature for an inert particle which has the value of the diffusion coefficient of the migrating cells, is very high $\approx 10^5 $ K ($r = 10 \mu m, \eta = 0.01 P$) \cite{franschi}. 

\begin{figure} 
\epsfxsize=7cm
\epsfbox{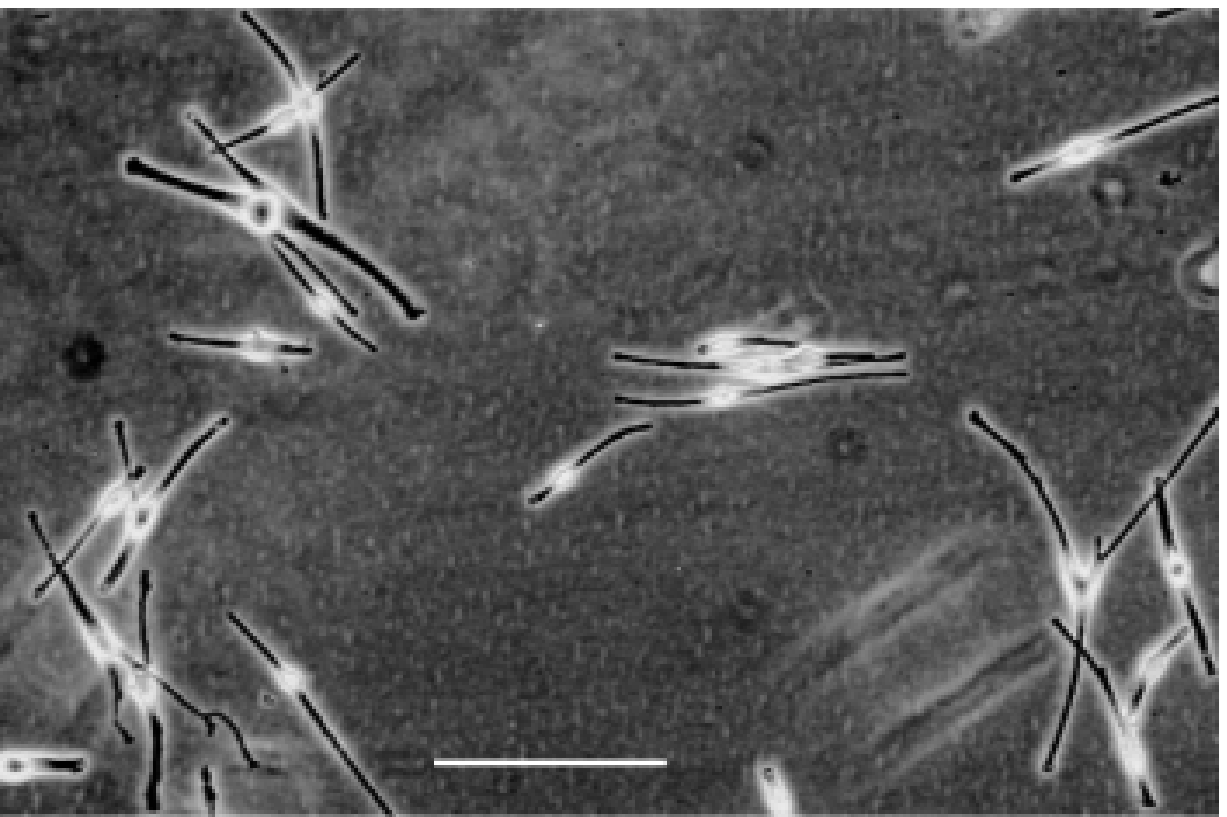}
\epsfxsize=7cm
\epsfbox{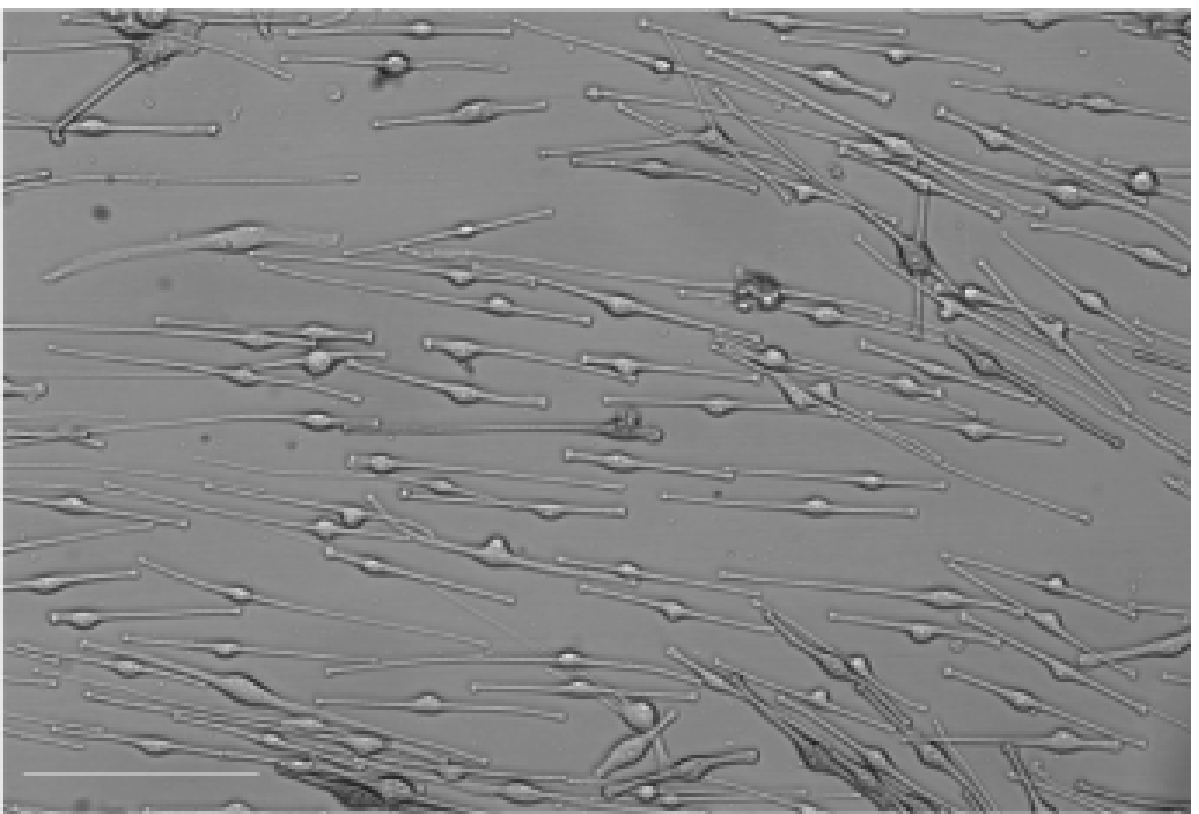}
\caption{\label{Melano21} \it {
Human melanocytes on a plastic surface. The bar represents 100 $\mu$m: a) Melanozytes spread at low cell density. Small sized clusters with oriented cells are formed. b) At high cell density the cell density becomes more uniform and the cells are oriented over large distances.  
}}
\end{figure}

(ii) An apolar nematic liquid crystal is expected if the cell-cell interaction is apolar. This means that one cell can influence the orientation of another cell. The cellular response is such that the cells like to be parallel. For example, human melanocytes are elongated cells which distribute with their dendrites the pigment melanin in the skin. Although the involved chemical signal is unknown there exists an attractive cell-cell interaction because the cell body of one cell attracts another cell body. There also exists a repulsive cell-cell interaction since the dendrites avoid contact with each other. The elongated cells try to be parallel oriented to each other and form at high cell density a nematic liquid crystal. At low cell density the cells form cluster with oriented cells as shown in Fig. \ref{Melano21}a, however, with increasing cell concentration the cells form a nematic liquid crystal. Fig. \ref{Melano21}b shows a nematic liquid crystal close to the nematic-isotropic transition.  This observation is not restricted to melanocytes. Other cell types like fibroblasts, osteoblasts, adipocytes  etc. orient in the same way so that, in the mean, the cells are parallel to each other. The cells form an apolar nematic liquid crystal because there is no order in the center of mass of the cells but order in the cell orientation. The apolar symmetry of this phase will be demonstrated by investigating orientational defects as shown below. A few words to the dynamic behavior: Time lapse movies of melanocytes show that the cell body performs rhythmic movements and, in addition, each cell extend and retract periodically its opposing dendrites. The dendrites find their orientation with minimum interaction with other cells \cite{teichgraeber}.

\begin{figure}
\epsfxsize=7cm
\epsfbox{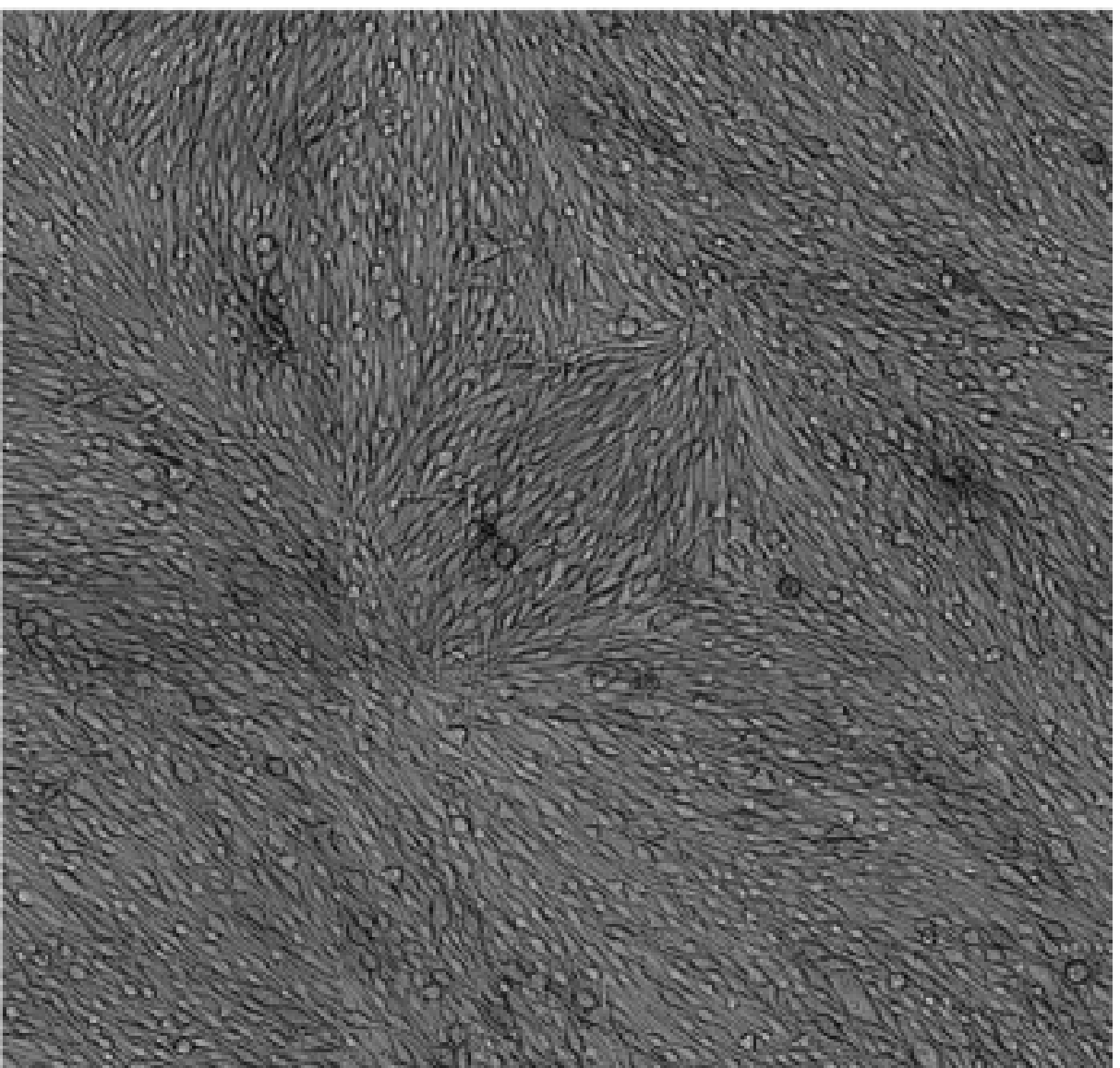}
\epsfxsize=7cm
\epsfbox{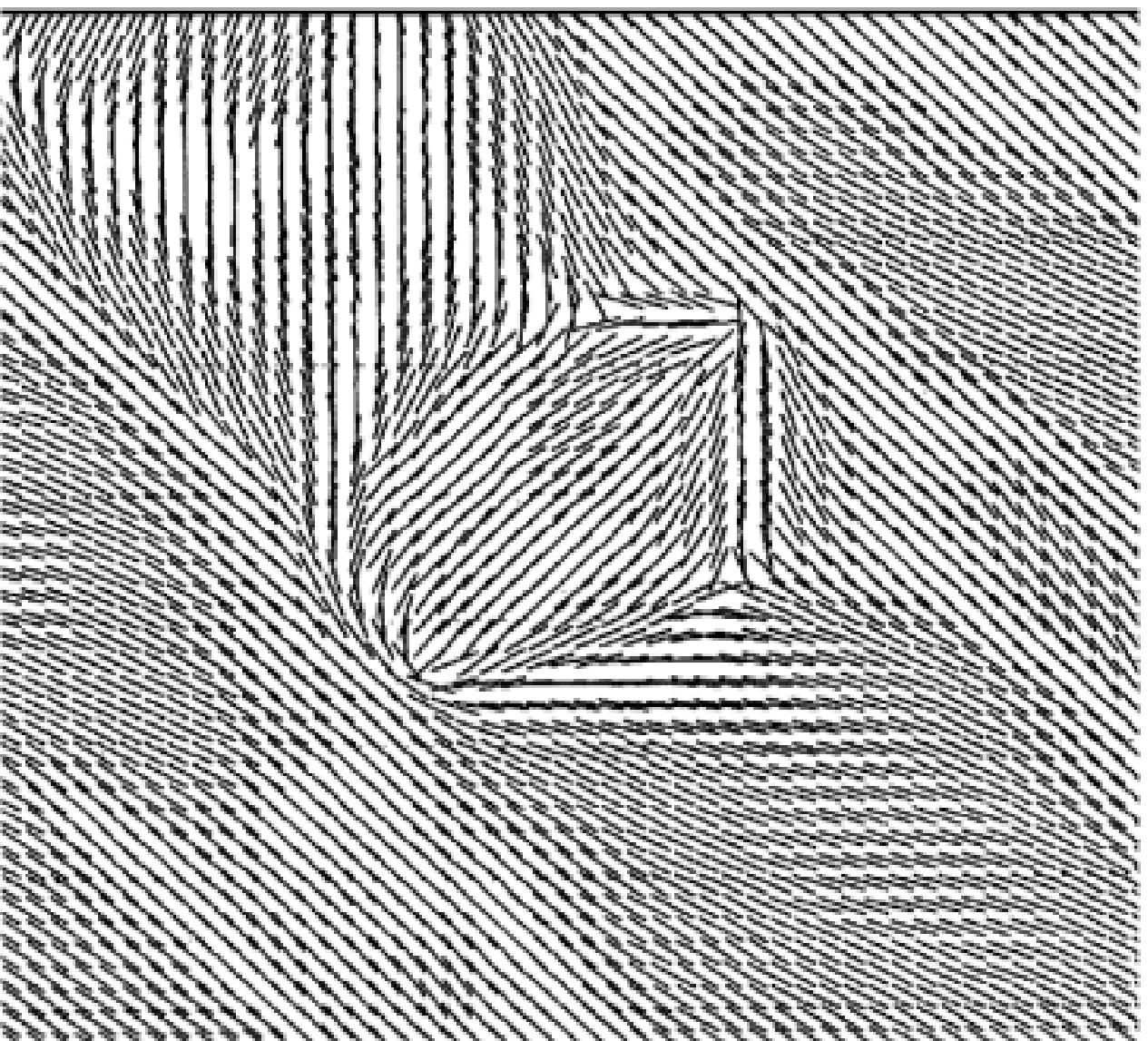}
\caption{\label{Melano41} \it {
a) Human melanocytes exposed to a smooth glass surface do not form an uniform apolar nematic liquid crystal. b) The calculated mean cell orientation of a) is shown.
}}
\end{figure}

A uniform oriented nematic liquid crystal is formed if a small extracellular guiding signal is applied like parallel scratches in the surface (Fig. \ref{Melano21}). With this method oriented single nematic liquid crystals were produced which had a size of several cm$^2$. An isolated single cell is not oriented by such a weak extracellular guiding signal \cite{dewald}. The uniform oriented nematic liquid crystals which are used in the liquid crystal display technology, are made in a very similar way \cite{degennes}. In the absence of an orienting extracellular guiding signal the nematic phase contains numerous orientational point defects (disclinations). An example is given in Fig. \ref{Melano41}. 
 
Summarizing this part, nematic liquid crystal phases can be formed either by anisotropic interacting building blocks at thermal equilibrium or by anisotropic interacting building blocks far from thermodynamic equilibrium. 

\section*{Orientational Elastic Energy}

\begin{figure}
\epsfxsize=7cm
\epsfbox{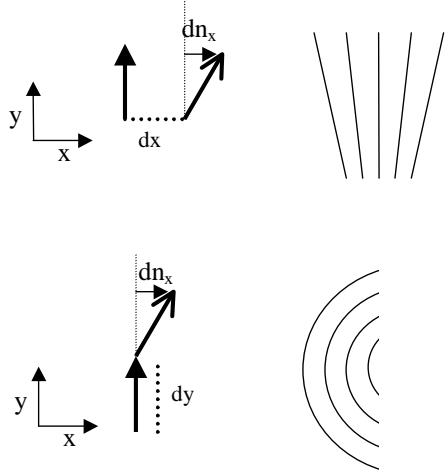}
\caption{\label{SplayBend} \it {
A splay deformation is shown on the top and a bend deformation on the bottom. On the left side the change of the director described by $\vec{n}$ is shown, if the space coordinate is changed by an increment $dx$ or $dy$. The flux line of the director is shown on the right side (spay on the top and bend on the bottom).
}}
\end{figure}

An apolar nematic liquid crystal is an uniaxial system ( $\pm \vec{n}$) where the molecules are on average aligned along the director  $\pm \vec{n}$. In the case of an ideal nematic liquid crystal, the direction of the director $\vec{n}$ is the same all over the sample.  However, this ideal conformation will not be compatible with the constrains that are imposed by limiting boundaries. There will be some deformation of the alignment. The distance $\ell$, over which significant variations of orientation occur, is of most interest, and is much larger than the dimensions, $a$, of the building blocks. Thus, the deformations may be described by a continuum theory disregarding the details of the structure on the scale of the building blocks ($a/\ell<<1$). To construct such a theory one possible starting point would be the elastic distortion energy density $F_d$ (\cite{oseen,frank}). The spatial derivatives of $\vec{n}(\vec{r})$ are used to construct $F_d$. They form a tensor of rank two $\partial_\alpha n_\beta$. At any point we introduce a local system of Cartesian co-ordinates, x, y, z with y parallel to $\vec{n}$ at the origin, x chosen perpendicular to y (x-y plane are the plane where the cells migrate) and z parallel to the normal of the plane. x, y, z form a right-handed system. Referred to these axis, the two components of curvature, at this point, are the splay and bend deformation $s = \partial n_x/\partial x$ and $b = \partial n_x/\partial y$, respectively (see Fig. \ref{SplayBend}). The distortion energy of a liquid crystal specimen in a particular configuration, relative to its energy in the state of uniform orientation, is expressible at the area integral of distortion energy density $g$ which is a quadratic function of two differential coefficients which measure the curvature in two dimensions. 
\begin{equation}
g = k_1s + k_3 b + \frac{1}{2} k_{11} s^2 + \frac{1}{2} k_{13}s b + \frac{1}{2} k_{33} b^2 + \cdots
\end{equation} 
The spontaneous splay coefficient, $k_1$ is zero if the axis perpendicular to $\vec{n}$ is a symmetry axis ($\leftarrow = \rightarrow$). The spontaneous bend coefficient, $k_3$, is zero if the axis parallel to $\vec{n}$ is a symmetry axis ($\uparrow = \downarrow$). The splay-bend coefficient, $k_{13}$, is only unequal to zero for $\rightarrow \neq \leftarrow$ and $\uparrow \neq \downarrow$. $k_{11}$ and $k_{33}$ are the orientational elastic coefficient for splay and bend deformation. These coefficients are also known as Frank elastic constants.
  
The distortion energy can be written in the coordinate free notation
\begin{equation}
g = k_1 \nabla \vec{n} + k_3 (curl \vec{n})_z + \frac{1}{2} k_{11} (\nabla \vec{n})^2 + \frac{1}{2} k_{13}(\nabla \vec{n})(curl \vec{n})_z + \frac{1}{2} k_{33} (curl \vec{n})_z^2
\end{equation}

\section*{Orientational Elastic Constants}

\subsection*{Splay and Bend Constants}

The orientational elastic constant of a nematic liquid crystal (bulk) can be estimated by purely dimensional arguments: one expects that the elastic constants are to be in the order of $U/a$, where $U$ is a typical interaction energy between the molecules while $a$ is a molecular dimension. A typical value of the molecular interaction energy is $\approx$ 8 kJ/mol or $1.3 \times 10^{-20}$ J/molecule which corresponds to a temperature of $\approx 2000$ K. A typical liquid crystal molecule has a length of about 1.5 nm. It leads to a value of $\approx 8 \times 10^{-12} $ N which is in accordance with the experiments \cite{degennes}. The orientational elastic constant of a membrane can be estimated by taking this value times the thickness, d, of the membrane. It yields $3 \times 10^{-20}$ J which is again verified by experiments \cite{helfrich}.

The orientational elastic constant of a nematic liquid crystal formed by living cells can be estimated in a similar way: In the case of migrating granulocytes the quasi temperature is $\approx 10^5$ K. A condensed phase can only be formed if the interaction potential, U, is in this order or even larger. This leads to an interaction energy of $\approx 5 \times 10^{-19}$ J/cell. The bulk elastic constant is obtained by dividing this interaction energy by the length of a cell ($\approx 20 \mu$m). It yields $2.5 \times 10^{-14}$ N. The two dimensional elastic constant is obtained by multiplying this value with the cell thickness ($\approx 20 \mu$m). The estimated orientational elastic constant of a quasi two dimensional nematic liquid crystal is then $\approx 5\times 10^{-19}$ J.

The dependence of the orientational elastic coefficients on the apolar order parameter, $S$, and the cell density, $\rho$, can be estimated in the following way: 
The orientation angle of amoeboid cells is controlled by an automatic controller which compares the actual angle with the desired one \cite{peliti}. The cellular response is in such a way as to minimize the deviation between the actual angle and the desired one. The desired angle of one cell is to be parallel to another cell. In the case of a nematic liquid crystal the information of the desired angle is transferred to one cell by an extracellular guiding field which can be chemical, electrical or sterical. The extracellular guiding field, $E_2$, of a nematic liquid crystal can be approached by the nematic mean field where one cell is considered in the field produced by all the other cells.  The strength of the mean field is assumed to be proportional to (i) the cell density, $\rho$, and (ii) the mean apolar order parameter, $S(=\langle cos 2 \Theta \rangle)$, of the surrounding cells. 
\begin{equation} \label{}
E_2 = a_2 \, \rho \, S
\end{equation}
These assumptions are verified by experiments \cite{dewald}. The extracellular guiding field is calibrated in cellular units by $a_2$.

The cellular automatic controller acts in such a way as to minimize the deviation between the actual angle and the desired one. In the absence of extracellular guiding field, the cells alter their angle of orientation in a random fashion. The response of the cellular automatic controller is hence not simply described by the extracellular guiding field but also by stochastic processes in the cellular signal transduction chain \cite{peliti}. The rate equation for the orientation angle is a stochastic differential equation (Langevin equation):
\begin{equation} \label{langevin1}
\frac{d \Theta}{d t} = - k_2 \, E_2 \, sin 2 \Theta + \Gamma_2(t)
\end{equation}
The orientation angle, $\Theta$, is measured in respect to the director $\vec{n}$ of the nematic liquid crystal. The cellular signal transformer is described by $E_2 sin 2 \Theta$ and the cellular reaction unit by the coefficient $k_2$. The stochastic part of the machine equation, $\Gamma(t)$, is approximated by a white noise of the strength $q_2$ ($\langle \Gamma \rangle =0, \, \langle \Gamma(t) \, \Gamma(t') \rangle = q_2 \, \delta(t-t')$) . The Langevin equation can be transformed into a Fokker-Planck equation which describes the temporal variations of the angle distribution function \cite{risken,franschi}. The predicted steady state angle distribution function is 
\begin{equation} \label{boltzmann1}
f_2(\Theta) = f_{20} \, e^{V \, cos 2 \, \Theta}
\end{equation}
with the generating function V
\begin{equation}
V = \frac{2 \, k_2 \, E_2}{q_2} = \frac{2 \, k_2 \, a_2 \, \rho \ S}{q_2} = A_2 \, \rho \, S
\end{equation}
$f_{20}$ is determined by the normalization ($\int f(\Theta) \, d \Theta = 1$). It is important to realize that the steady state angle distribution function of living cells (system far from thermodynamic equilibrium) have the same mathematical structure as the Boltzmann distribution which describes the fluctuations of a system at thermal equilibrium. The predicted angle distribution function, $f_2(\Theta)$, of a nematic liquid crystal formed by migrating and orienting cells as well as the density- and order parameter dependence of the extracellular guiding field are verified by experiments \cite{dewald}. The unknown coefficient, $A_2 \, (=2 \, k_2 \, a_2 \, / \, q_2)$, can be determined by measuring the angle distribution function at different cell densities. For melanocytes one gets $1/A_2$ = 55 cells/mm$^2$.  Form the self-consistency for the order parameter one gets the transition from the nematic to the  the isotropic phase \cite{dewald}: $A_2 \, \rho_0 = 2$ with cell density, $\rho_0$, at the transition. This threshold cell density, $\rho_0$, can be estimated by considering random oriented cells without steric contact. One gets a threshold cell density of $\approx 100$ cells/mm$^2$ for  $\approx 100 \mu$m long cells. This value is in accordance with $\rho_0$ (= 110 cells/mm$^2$) determined directly from $A_2$. 

The interaction potential, $U$, of a single cell with its environment is the generating function, $V$, times $\cos 2 \, \Theta$
\begin{equation} \label{IntPot}
U = A_2 \, \rho \, S \, \cos 2 \, \Theta = 2 \frac{\rho}{\rho_0} \, S \, \cos 2 \Theta
\end{equation}
This interaction potential of a single cell in the environment of a nematic liquid crystal will be used to predict the order parameter dependence as well as the density dependence of the orientational elastic coefficients. The calculation is made in analogy to the elastic coefficients of nematic liquid crystal formed by molecules \cite{gruler74}.

The angle $\Theta$ alters by the distortion angle $\alpha$ if one moves in space (y-direction). One obtains
\begin{equation} \label{alpdaDis1}
\Theta \rightarrow  \Theta +  \alpha
\end{equation}
For small  distortion angle one gets
\begin{equation} \label{alphaDis2}
\cos2 \,(\Theta +  \alpha) =  (1 - 2 \, \alpha^2) \, \cos 2 \, \Theta  -  2 \, \alpha \, \sin 2 \, \Theta 
\end{equation}
The mean interaction potential per area is obtained from Eqs. \ref{IntPot} - \ref{alphaDis2} as
\begin{equation}
u = 2 \, \frac{\rho^2}{\rho_0} \, S^2 \, (1 - 2 \, \alpha^2)
\end{equation}
In the case of a bend deformation, the distortion angle is on a molecular scale 
\begin{equation}
\alpha = \frac{\partial \Theta}{\partial y} \xi_y
\end{equation}
$\xi_y$ represents the characteristic length of a bend deformation. One expects that $\xi_y$ is the length of one elongated cell. The elastic distortion area density is then
\begin{equation}
u_{33} = 4 \, S^2 \, \frac{ \rho^2}{\rho_0} \, \xi_y^2  \, \left( \frac{\partial \Theta}{\partial y} \right)^2 
\end{equation}
which has to be compared with the phenomenological derived bend distortion energy 
\begin{equation}
g_{33} = \frac{1}{2} \, k_{33} \, \left( \frac{\partial \Theta}{\partial y} \right)^2
\end{equation}
The splay elastic constant is then
\begin{equation} \label{k_mean}
k_{33} = 8 \, b \, \frac{\rho^2}{\rho_0}(S \, \xi_y)^2
\end{equation}
A similar expression can be derived for the splay elastic constant, $k_{11}$. One expects that the elastic coefficient increases with (i) increasing order parameter, (ii) increasing cell density and (iii) increasing cell dimension. A factor $\approx 500$ is expected between the value at the nematic-isotropic transition ($\rho_0 = 110$ cells/mm$^2$ and $S \approx 0.5$) and that of a nematic liquid crystal film of dense packed cells ($\rho \approx 10 \, \rho_0$ and $S \approx 1$). 

The ratio of the elastic constants is determined by the ratio of characteristic length of the splay and bend deformation.
\begin{equation}
\frac{k_{11}}{k_{33}} = \left( \frac{\xi_x}{\xi_y} \right)^2
\end{equation}
One expects that these characteristic lengths, $\xi_x$ and $\xi_y$, are approximately $ a$. Only small differences between  $\xi_x$ and $\xi_y$ are expected since a cluster of cells will change their orientation if one tries to change the orientation of a single cell. 

The coefficient $b$ transforms the cell specific physical units into man-made units. Up to now no orientational elastic measurements were performed and consequently  the coefficient, $b$, is unknown. It can be estimated in the following way: The orientational elastic coefficient close to the nematic-isotropic transition is expected to be the interaction potential, $U$, divided by the cellular length, $a$, and multiplied by the thickness, $d$, of the nematic film ($U \cdot d/a$) as shown above. The orientational elastic coefficient close to the transition is predicted by Eq. \ref{k_mean} to $2 \, b \, \rho_0 \, \xi_y^2$ (with $S \approx 0.5$ at the transition). This leads to 
\begin{equation}
b \approx \frac{1}{2} \, \frac{d }{\rho_0 \, a^3} \, U
\end{equation}
One gets $b \approx 5 \times 10^{-20}$ J if one uses $U \approx 5 \times 10^{-19}$ J/cell as estimated above and $d \approx 20 \mu$m,  $a \approx 100 \mu$m, $\rho_0 = 110$ cells/mm$^2$. 

\subsection*{Spontaneous Splay}

The spontaneous splay coefficient, $k_1$ is zero if the axis perpendicular to $\vec{n}$ is a symmetry axis. If the building blocks carry a preferred direction in the direction of the long axis (e.g. wedge-shaped building blocks) there are as many building blocks up ($\uparrow$) as there are down ($\downarrow$) then the spontaneous splay coefficient is zero. A (parallel) polar nematic liquid crystal with its (parallel) polar distribution function ($\uparrow \neq \downarrow$) is formed in case of a (parallel) polar cell-cell interaction. The (parallel) polar distribution and the (parallel) polar shape asymmetry of the building blocks leads to a natural splay of the structure and the spontaneous splay coefficient, $k_1$ is expected to be unequal to zero. If the (parallel) polar anisotropy of the building-blocks (e.g. wedge-shaped cell, (parallel) polar organized cell structure) is connected with a (parallel) polar distribution of the muscle proteins then the center of mass of an area element moves in the direction of the director field.  In case of a point defect $div \vec{v} \ne 0$.

To proceed further one has to introduce the motor characteristics of the migrating cells. The excess force, $F(c, v)$, of a migrating cell depends on the concentration, $c$, of the migration stimulating molecules in the extracellular space and on the speed, $v$, of the cell \cite{lelievre}. 
\begin{equation} \label{}
F(c, v) = F_0 \, \frac{c}{c + K_c} (1 - \frac{v}{v_{max}} )
\end{equation}
The first term describes the maximum force, $F_0$, which is 40 nN for a granulocyte. The second term describes the fraction of membrane-bound receptors loaded with the migration stimulating  molecule. The maximum speed, $v_{max}$, is 24 $\mu$m/min for granulocytes. One expects movements as demonstrated in Fig. \ref{Cluster1}b. The hydrodynamic equations for nematic liquid crystals formed by migrating cells are not yet developed. 

\subsection*{Spontaneous Bend}

The spontaneous bend coefficient, $k_3$, is zero if the axis parallel to $\vec{n}$ is a symmetry axis. If the building blocks carry a preferred direction perpendicular to the direction of the long axis (e.g. banana-shaped building blocks or elongated cells with the leading front on one side (e.g. keratinocytes)), there are as many building blocks right ($\rightarrow$) as there are left ($\leftarrow$) then the spontaneous bend coefficient is zero. A (transverse) polar nematic liquid crystal with its (transverse) polar distribution function ($\uparrow \neq \downarrow$) is formed in case of a (transverse) polar cell-cell interaction. The (transverse) polar distribution and the (transverse) polar shape asymmetry of the building blocks leads to a natural bend of the structure and the spontaneous bend coefficient, $k_3$ is expected to be unequal to zero. Human keratinocytes (skin cells) plated on a surface migrate and cluster in large islands. The elongated keratinocytes in the cluster show the typical paving stone appearance \cite{hegemann}. If the (transverse) polar anisotropy of the building-blocks (e.g. banana-shaped cell, (transverse) polar organized cell structure) is connected with a (transverse) polar distribution of the muscle proteins then the center of mass of an area element moves perpendicular to the director field. In case of a point defect one expects $curl_z \vec{v} \ne 0$.

\section*{Disclinations}

The basic assumption of the used continuum theory concerns the smoothness of the director field, $\vec{n}(\vec{r})$. In practice, however, textures are observed originating from singularities in the orientational field. These discontinuities in the orientation are called disclinations. The energy of the disclinations themselves is unknown and the energy of the surrounding distorted director field can be calculated by making use of the elastic continuum theory.

The director is given by
\begin{equation}
\vec{n}(\vec{r}) = [\cos \Phi(x,y), \sin \Phi(x,y)]
\end{equation}
where $\Phi (x,y)$ is the angle between the director and the x-axis of a fixed coordinate system. In order to avoid complicated mathematics the one-constant approximation is used $k = k_{11} = k_{33}$.
\begin{equation}
g = \frac{1}{2} \, k \, (div \, \vec{n})^2
\end{equation}
The cylindrical coordinates ($r, \psi $) are used to describe the singularity.
\begin{equation}
g(r, \psi) = \frac{1}{2} \, k \,\left[ \left( \frac{\partial \Phi}{\partial r}\right)^2 + \frac{1}{r^2} \, \left( \frac{\partial \Phi}{\partial \psi}\right)^2 \right]
\end{equation}
It is evident that this expression diverges at $r = 0$. The cause is a discontinuity in orientation at $r = 0$.

\begin{figure}
\epsfxsize=7cm
\epsfbox{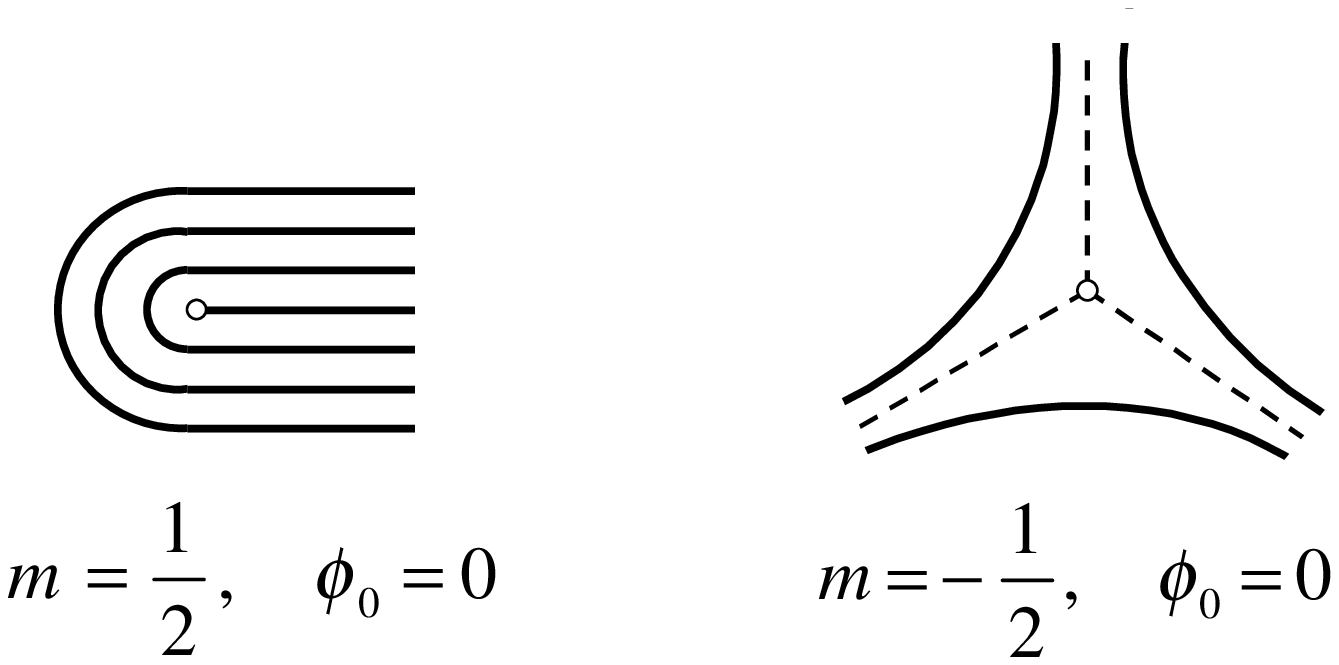}
\epsfxsize=7cm
\epsfbox{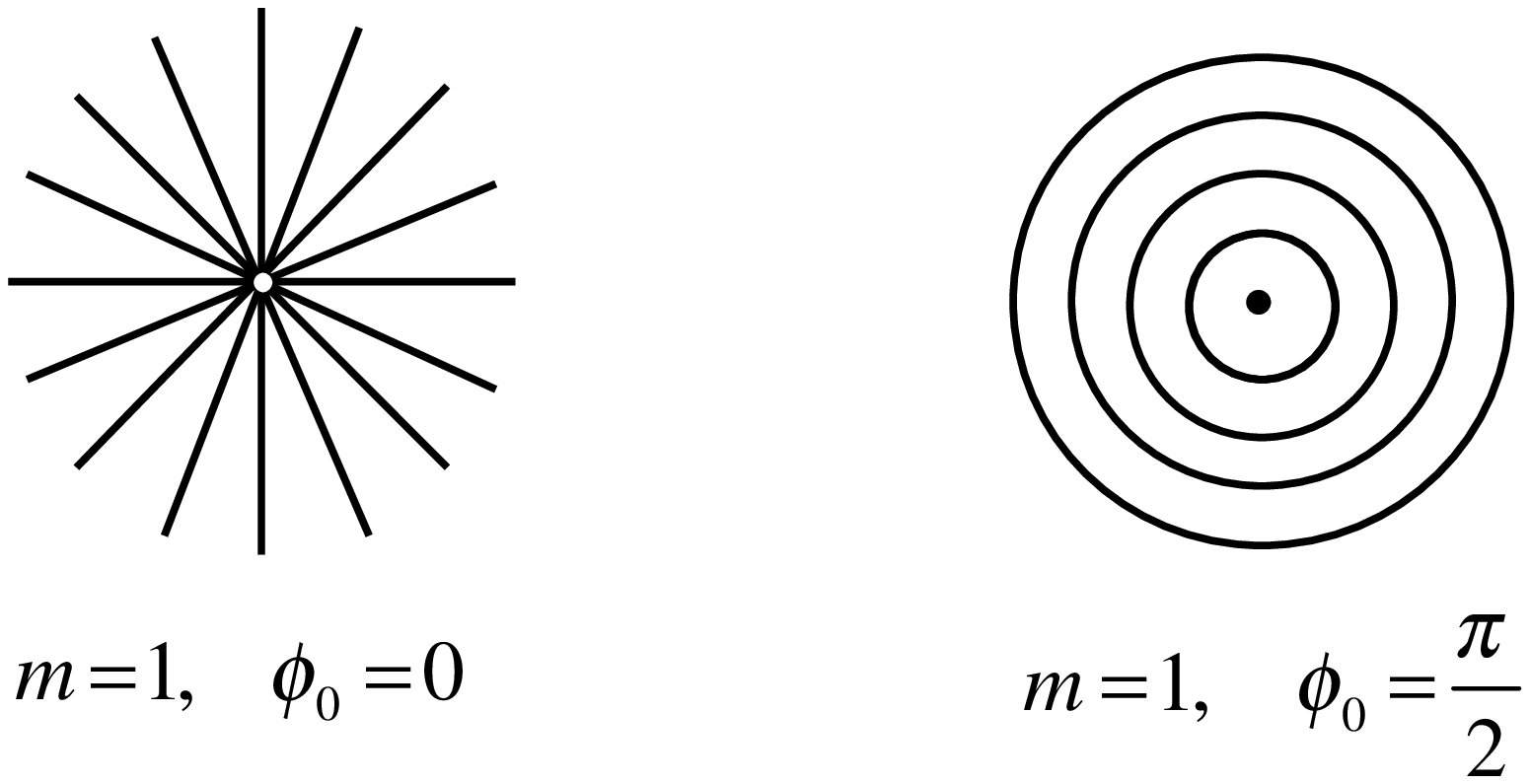}
\caption{\label{DisHalbGanz} \it {
A schematic drawing of the flux lines of the director field is shown for four possible disclinations: a) and b) shows the diclinations for $m = \pm 0.5$ and $\Phi_0 = 0$ and c) and d) the disclinations for $m = 1$ and $\Phi_0 = 0$ and $\pi/2$. }}
\end{figure}

The functional dependence of $\Phi$ on $r$ and $\psi$ is determined by requiring that the distortion energy must be minimum with respect to variations of $\Phi$. This means that $\Phi$ must satisfy the Euler-Lagrange equation.
\begin{equation}
\frac{\partial^2 \Phi}{\partial r^2} + \frac{1}{r} \, \frac{\partial \Phi}{\partial r} + \frac{1}{r^2} \, \frac{\partial^2 \Phi}{\partial \psi^2} = 0
\end{equation}
The trivial solution is $\Phi = const$, where the cells form an uniform oriented nematic liquid crystal. The general solution for the disclinations is
\begin{equation}
\Phi = m \, \psi + \Phi_0
\end{equation}
For an apolar nematic liquid crystal 2m is an integer and for a polar nematic liquid crystal m is an integer. Some examples are shown in Fig. \ref{DisHalbGanz}. 

\begin{figure}
\epsfxsize=7cm
\epsfbox{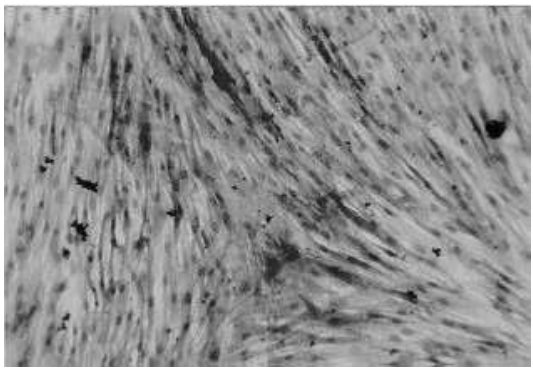}
\caption{\label{Osteoblast} \it {A disclination ($m = - 0.5$) is shown for osteoblasts. }}
\end{figure}

The question whether the nematic liquid crystal formed by living and migrating cells has a polar or an apolar structure can be decided by investigating the types of defects. (i) Dense packed elongated melanocytes form an apolar nematic liquid crystal since the half-number disclination ($m = - 1/2$) is found very often in the orientation pattern (Fig. \ref{Melano41}). However, this result is not specific for melanocytes. Other cell types like fibroblasts \cite{goldman}, osteoblasts (Fig. \ref{Osteoblast}), etc. show half-number disclination and, thus, form an apolar nematic liquid crystal. (ii) No half-numbered defects are found in the dense packed clusters of migrating granulocytes and keratinocytes. Granulocytes form a (parallel) polar nematic phase since this cell type like to create a defect with $m = 1$ and $\Phi_0 = 0$ (Fig. \ref{Cluster1}).  Keratinocytes form a (transverse) polar nematic phase since this cell type like to create a defect with $m = 1$ and $\Phi_0 = \pi /2$. 

\section*{Interacting Disclinations}

The deformation energy, W, of an isolated disclination can be obtained by integrating the elastic energy density in a circular area around the disclination.
\begin{equation}
W = W_c + \pi \, k \, m^2 \, \ln \left( \frac{R}{R_c} \right) 
\end{equation}
R, $R_c$ and $W_c$ are the radius of the circular integration area, the core radius of the disclination and the core energy of the disclination, respectively. The elastic energy, $W - W_c$, depends on the type of disclination. The elastic energy of a $\pm 0.5$-disclination is a forth of that of a $\pm 1$-disclination. Therefore, the $\pm 0.5$-disclination should occur more frequently than the $\pm 1$-disclination. This prediction is in accordance with the observation. 

The deformation field of one disclination is altered if another disclination is introduced. The elastic interaction energy, $W_{12}$, is then \cite{degennes}
\begin{equation}
W_{12} = 2  k  \pi  m_1 m_2 \ln \left(\frac{R_{12}}{R_c} \right)
\end{equation}
where $R_{12}$ is the distance between the two disclinations. As expected the interaction between disclination of opposite strength like $m_1 = +0.5$ and $m_2 = - 0.5$ is predicted to be attractive  and repulsive for equal signs. The force between the two disclination is \cite{degennes}
\begin{equation}
F_{12} = - \frac{2 \, k \, \pi \,  m_1 \, m_2}{R_{12}}
\end{equation}
As expected the force is predicted to be large for small distances and small for large distances.

\begin{figure}
\epsfxsize=7cm
\epsfbox{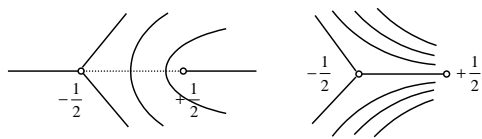}
\caption{\label{interact} \it {Two possible orientations of a +0.5-disclination and -0.5-disclination are shown. The left side is preferred for $k_{11}>k_{33}$ where the bend deformation is preferred and the ride side for $k_{11}<k_{33}$ where the splay deformation is preferred.}}
\end{figure}

The elastic interaction between two disclinations is orientation dependent: Two possible orientations in the case of a $-0.5$ disclination and a $+0.5$ disclination where the interaction energy is minimum, are shown in Fig. \ref{interact}. The elastic interaction energy of the two orientations is the same in the case of equal elastic constants ($k = k_{11} = k_{33}$) is used. But, in the general case ($k_{11} \neq k_{33}$) the elastic interaction energy is minimum for one of the orientations \cite{russe}.  The orientation as shown in Fig. \ref{interact}a is energetically preferred in the case of $k_{11} > k_{33}$ since a lot of bend deformation is between the disclinations. The orientation as shown in Fig. \ref{interact}b is energetically preferred in the case of $k_{11} < k_{33}$ since a lot of splay deformation is between the disclinations. The splay elastic constant, $k_{11}$, is smaller than the bend elastic constant, $k_{33}$,  for an apolar nematic liquid crystal formed by melanocytes since in the case of a pair of $\pm 0.5$ disclination the orientation as shown in Fig. \ref{interact}b is preferentially formed. But up to now, we are not able to determine the relative value of the anisotropy, $2 (k_{11} - k_{33})/(k_{11} + k_{33})$, of the elastic constants.

\section*{Core of a Disclination}

\begin{figure}
\epsfxsize=10cm
\epsfbox{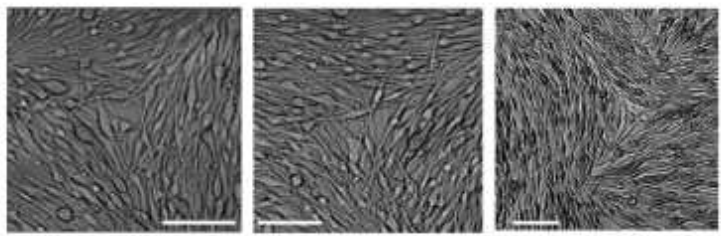}
\caption{\label{DisHalbEx} \it {
A $m = -0.5$ disclination is shown for melanocytes. The bar is 100 $\mu$m. Different possibilites are shown: a) The core of the disclination is an area free of cells. b) The core of the disclination is an area with isotropically distributed cell. c) The core of the disclination is occupied by a star-shaped cell. The cells which form the nematic liquid crystal are in an elongated bipolar state.
}}
\end{figure}

The nature of the core of nematic liquid crystals formed by anisotropic molecules is not known. However, the core of a nematic liquid crystal formed by migrating cells can be observed directly in the light microscope. Different types of cores are observed as shown in Fig. \ref{DisHalbEx}: (i) The core of a disclination is an area free of cells. This type of core is frequently observed for very elongated cells like melanocytes (Fig. \ref{DisHalbEx}a). (ii) The core of a disclination is an area with isotropically distributed cell as shown in Fig. \ref{DisHalbEx}b. This type of core is frequently observed for smaller cells like fibroblasts and osteoblasts. (iii) The core of a disclination is occupied by a star-shaped cell as shown in Fig. \ref{DisHalbEx}c. The cells which form the nematic liquid crystal are in an elongated bipolar state.

The defect in the orientation field attracts disturbances of the orientation field. One observes that other cell types or the same cell type with different shapes are trapped in the core of a disclination. For example, melanocytes have usual two opposing dendrites. But some cells have three dendrites and these cells disturb the nematic phase and are, therefore, attracted by the orientational disclination (Fig. \ref{DisHalbEx}c).

In nematic liquid crystals formed by elongated molecules it is assumed that the molecules in the core of a disclination are in the isotropic state. This assumption can be verified for a nematic liquid crystal formed by elongated migrating cells.  There exist a sharp boundary between the nematic and the isotropic phase. The nematic phase is destroyed if the elastic energy density is larger than the mean field.
\begin{equation}
\frac{1}{2} \, k_{33} \left( \frac{\partial \Theta}{\partial y} \right) ^2 \ge 2 \, b \, \frac{(\rho \, S)^2}{\rho_0} \,  
\end{equation}
The bend elastic constant divided by the calibration coefficient b can be expressed by Eq. \ref{}. One gets
\begin{equation} \label{disclinCore}
\frac{\partial \Theta}{\partial y}  \ge \frac{1}{\sqrt{2}} \, \xi_y
\end{equation}
The maximal bending $\partial \Theta / \partial y$ of a +1/2 disclination can be determined from photographic pictures. The characteristic length, $\xi_y$, of a bend deformation is expected to be in the order of the length of an elongated cell ($\xi_y \approx a$). The observed angular change by going from one cell to the next is approximately 15 to 20 degrees. The angular change per cell as predicted by Eq. \ref{disclinCore} is 29 degrees. The assumption that a strong bend deformation can destroy the nematic phase is verified since the predicted angular change is in accordance with the experimental observation.

\section*{Outlook}

The practical importance of this study lies in its objective description of the liquid crystal phases formed by migrating and interacting cells. Two types of nematic liquid crystals, a polar and an apolar phase, are found to be formed by certain cell types in culture and such phases may be expected to be present in many other cell types. In addition, the objective description may be useful to define and elucidate cases of cellular dysfunction, and of altered cellular function induced by specific molecules like pharmacological active molecules, etc..

The major objective of current and future research is to go beyond this phenomenological description of the thermodynamic phases formed by migrating and interacting cells. How the cellular machinery works on a molecular scale can be investigated by considering the technical data as a frame filled with the biochemical events. The cell, regarded as a fluid self-organized molecular machine, represents a new and important field in the triangle between physics, chemistry and life sciences. The knowledge that one essential process in the cellular signal transduction chain is the supply of fresh receptors to the plasma membrane could lead to the application of new techniques. For instance the liquid crystal phase can be changed if special designed molecules are added which interact with the intracellular signal transduction chain and alter cell-cell interaction either by the induced change in the cellular shape or by the transmitted chemical signal.

The process by which migrating cells exchange information constitutes one of the most intriguing areas of life sciences and physics. An individual cell transmits signals and guides its migration and orientation according to the signals it receives, these messages being chemical, sterical or electrical in nature. The physical process of reducing a gas containing freely moving molecules to a liquid form is understood to a large extent and the mean behavior of a given molecule in the interaction field of the other moving molecules may be determined by Boltzmann statistics. Identical principle are used to reduce freely migrating cells to a condensed state. The mean behavior of a given cell in the interacting field of other cells can be described in terms of its automatic controller. The self-organization, involved in morphogenesis, organogenesis and wound healing, takes in a new light.

The nematic liquid crystal formed by migrating cells opens new perspectives in the field of cellular tissue engineering. The two dimensional nematic liquid crystal can be used as a template. For example: (i) Cells of another type can migrate and orient on top of the liquid crystal to form more complex structures. (ii) Cells of another type can be attracted and localized in the orientational defects of the nematic liquid crystal. New types of liquid crystals are expected in analogy to blue phases where orientation defects are ordered in space. (iii) Complex structures formed by different types of cells can be altered if the phase of the template (cellular layer of the nematic liquid crystal) is altered by a chemical signal from a nematic liquid crystal phase to an isotropic phase.

\subsubsection*{Acknowledgments}
We particularly like to thank Volker Teichgr\"aber for fruitful discussion. This work was supported by Deutsche Forschungsgemeinschaft and Fonds der Chemischen Industrie.

\section*{Appendix}

\subsection*{Cell Preparation}

Melanocytes are one type of cells in the skin. They produce and distribute the color pigment melanin via dendrites. Cultured human melanocytes form the desired bipolar mode with two opposing dendrites when they are exposed to melanocytes medium. Details are given in ref.  \cite{kaufmann}.  Briefly, skin biopsies were obtained from healthy donors. The cells were cultured in TIC medium (Ham F10, 16 \% serum, 85nM PMA, 0.1 mM IBMX, 2.5 nM cholera toxin). Cells designated for the experiment were seeded in $25-cm^2$ dishes (Nunclon, Nunc, Germany) and cultured until confluence (approx. 25-30 days). 

The cells, being of spherical shape in the suspension, come down onto the substrate and start to develop their dendrites. They stay in this shape and crawl around on the surface. The cell body of different melanocytes attract each other but the dendrites repel each other. The orientation of the dendrites occurs during a periodic retraction and elongation process in such a way that each dendrite has a minimum of contact with the surrounding cells. 

We used a precision motorized x-y stage on a CCD-camera (Hamamatsu) equipped microscope (Zeiss Axiovert) and scanned an area of several mm$^2$ using a 10x magnification lens. The pictures were digitized with a Scion frame grabber card on a Macintosh PPC and stored on the hard disk. The public domain software NIH-IMAGE (developed at the US National Institute of Health) was used for this purpose, as well as, for the further image processing. 

\subsection*{Picture Evaluation}

\begin{figure}
\epsfxsize=7cm
\epsfbox{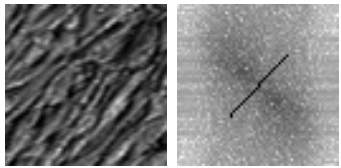}
\caption{\label{kling} \it {
A small rectangular frame of $64 \times 64$ pixels was chosen in the real space (photographic picture on the left side). The 2d-Fourier transformed of this picture is shown on the right side. The mean cell orientation is calculated by means of the Fourier Transformed and is shown by the bar. 
}}
\end{figure}

Our goal is to find an algorithm which determines automatically the mean orientation of the elongated cells in a small part of the picture. This task was successfully solved by going into the Fourier space. First, a squared section (64X64 pixels) of a picture is chosen, as shown in Fig. \ref{kling}a. A two-dimensional Fast Fourier Transformation implemented in NIH-IMAGE is applied to this squared section of the picture. The result is a gray scale picture as shown in Fig. \ref{kling}b. 

The Fourier transform arises as a diffuse cloud with a handle-shaped contour. The maximum momentum of this elongated cloud is determined and found to be parallel to the direction of the mean cell alignment.  In order to evaluate the exact orientation we apply a threshold procedure to remove all pixels with a gray value below a certainthreshold value. The remaining pixels are weighted with their gray value, g(x,y), and the angle, $\Theta$, of the orientation axis is found by maximizing the "momentum of mass". More details to the procedure are given in \cite{russ}.
\begin{equation}
tan \Theta = \frac{M_{xx} - M_{yy} + \sqrt{(M_{xx}-M_{yy})^2 +4 M_{xy}^2}}{2M_{xy}}
\end{equation}
with
\begin{equation}
M_{xx} = \sum x^2 g(x,y) - \left(\sum x g(x,y) \right)^2
\end{equation}
and M$_{xy}$ and M$_{yy}$ analogous. 

This procedure is repeated for the next squared section of the picture. The scan was performed with overlapping squared sections. The shift length was half of the square size.



\section*{Figures}

\newpage

\newpage

\newpage

\newpage

\newpage

\newpage

\newpage

\end{document}